\documentclass[12pt, a4paper]{article}

\usepackage{graphicx}

\begin{document}

\title{On Einstein's Doctoral Thesis
\footnote{Talk given at the joint colloquium of ETH and the University of Z\"{u}rich, 27
April (2005).}}

\author{Norbert Straumann\\
        Institute for Theoretical Physics University of Zurich,\\
        CH--8057 Zurich, Switzerland}

\maketitle

\begin{abstract}
Einstein's thesis ``A New Determination of Molecular Dimensions'' was the second of his
five celebrated papers in 1905. Although it is -- thanks to its widespread practical
applications -- the most quoted of his papers, it is less known than the other four. The
main aim of the talk is to show what exactly Einstein did in his dissertation. As an
important application of the theoretical results for the viscosity and diffusion of
solutions, he got (after eliminating a calculational error) an excellent value for the
Avogadro number from data for sugar dissolved in water. This was in agreement with the
value he and Planck had obtained from the black-body radiation. Two weeks after he finished
the `Doktorarbeit', Einstein submitted his paper on Brownian motion, in which the diffusion
formula of his thesis plays a crucial role.
\end{abstract}

\section{Introduction}

When Einstein's great papers of 1905 appeared in print, he was not a newcomer in the
\textit{Annalen der Physik}, where he published most of his early work. Of crucial
importance for his further research were three papers on the foundations of statistical
mechanics, in which he tried to fill what he considered to be a gap in the mechanical
foundations of thermodynamics. At the time when Einstein wrote his three papers he was not
familiar with the work of Gibbs and only partially with that of Boltzmann. Einstein's
papers form a bridge, parallel to the \textit{Elementary Principles of Statistical
Mechanics} by Gibbs in 1902, between Boltzmann's work and the modern approach to
statistical mechanics. In particular, Einstein independently formulated the distinction
between the microcanonical and canonical ensembles and derived the equilibrium distribution
for the canonical ensemble from the microcanonical distribution. Of special importance for
his later research was the derivation of the energy fluctuation formula for the canonical
ensemble.

Einstein's profound insight into the nature and size of fluctuations played a decisive role
for his most revolutionary contribution to physics: the light-quantum hypothesis. In this
first paper of 1905 he extracted the light-quantum postulate from a statistical mechanical
analogy between radiation in the Wien regime and a classical ideal gas of material
particles. In this consideration Boltzmann's principle, relating entropy and probability of
macroscopic states, played a key role. Later Einstein extended these considerations to an
analysis of fluctuations in the energy and momentum of the radiation field. For the latter
he was also drawing on ideas and methods he had developed in the course of his work on
Brownian motion, another beautiful application of fluctuation theory. This definitely
established the reality of atoms and molecules, and, more generally, gave strong support
for the molecular-kinetic theory of thermodynamics.

Einstein's Doctoral thesis ``A New Determination of Molecular Dimensions'' was the second
of his celebrated five papers in 1905. Unfortunately, it is not sufficiently well known.
The main body of the paper is devoted to the hydrodynamic derivation of a relation between
the coefficients of viscosity of a liquid with and without suspended particles. In
addition, Einstein derived a novel formula for the diffusion constant $D$ of suspended
microscopic particles. This was obtained on the basis of thermal and dynamical equilibrium
conditions, making use of van't Hoff's law for the osmotic pressure and Stokes' law for the
mobility of a particle. Einstein then applied these two relations to sugar dissolved in
water. Using empirical data he got (after eliminating a calculational error) an excellent
value of the Avogadro number and an estimate of the size of sugar molecules. Einstein's
thesis is the most quoted among his papers.

Soon afterwards Einstein's diffusion formula became also important in his work on Brownian
motion. In this celebrated paper he first gave a statistical mechanical foundation of the
osmotic pressure, and then repeated his earlier derivation in the thesis.

\section{Biographical remarks}

Einstein devoted his thesis to his friend Marcel Grossmann. Before I come to a technical
discussion of the paper, I would like to give some biographical and other background.

Until 1909 the ETH was not authorized to grant doctoral degrees. For this reason a special
arrangement enabled ETH students to obtain doctorates from the University. At the time most
dissertations in physics by ETH students were carried out under the supervision of H.F.
Weber, Einstein's former teacher at the `Polytechnikum' as it was then called. The
University of Z\"{u}rich had only one physics chair, held by Alfred Kleiner. His main
research was focused on measuring instruments, but he had an interest in the foundations of
physics. From letters to Mileva one can see that Einstein often had discussions with
Kleiner on a wide range of topics. Einstein also showed him his first dissertation in
November 1901. This dissertation has not survived, and it is not really clear what it
contained. At any rate, Einstein withdrew his dissertation in February 1902. One year later
he was giving up his plan to obtain a doctorate. To Besso he wrote: ``the whole comedy has
become tiresome for me''.

By March 1903 he seems to have changed his mind. Indeed, a letter to Besso contains some of
the central ideas of the 1905 dissertation, especially in the second part of the following
quote:
\begin{quote}
``\textit{Have you already calculated the absolute magnitude of ions on the assumption that
they are spheres and so large that the hydrodynamical equations for viscous fluids are
applicable? With our knowledge of the absolute magnitude of the electron [charge] this
would be a simple matter indeed. I would have done it myself but lack the reference
material and the time; you could also bring in diffusion in order to obtain information
about neutral salt molecules in solution.}''
\end{quote}

Kleiner was, of course, one of the two faculty reviewers of the dissertation, submitted by
Einstein to the University on 20 July, 1905. His judgement was very positive: ``the
arguments and calculations to be carried out are among the most difficult in
hydrodynamics''. The other reviewer, Heinrich Burkhardt, Professor for Mathematics at the
University, added: ``the mode of treatment demonstrates \textit{fundamental mastery of the
relevant mathematical methods}.''

In his biography of Einstein, Carl Seelig reports: ``Einstein later laughingly recounted
that his dissertation was first returned by Kleiner with the comment that it was too short.
After he had added a single sentence, it was accepted without further comment.''

The physical reality of atoms was not yet universally accepted by the end of the nineteenth
century. Fervent opponents were Wilhelm Ostwald and Georg Helm (who called themselves
``energeticists''), and Ernst Mach admitted only that atomism may have a heuristic or
didactic utility.

In his first three papers of 1905, Einstein found three different methods of determining
the Avogadro number. (A few years later he found another one in his study of critical
opalescence.) For him this was not only important for establishing the existence of atoms.
He later wrote to Perrin:
\begin{quote}
``\textit{A precise determination of the size of molecules seems to me of the highest
importance because Planck's radiation formula can be tested more precisely through such a
determination than through measurements on radiation.}''
\end{quote}

\section{Einstein's dissertation}

By 1905 several methods for determining molecular sizes were developed. The most reliable
ones were based on kinetic theory of gases. An important early example is Loschmidt's work
from 1865.\footnote{J. Loschmidt, Wiener Ber. \textbf{52}, 395 (1866). See also J.C.
Maxwell, \textit{Collected Works}, Vol. 2, p. 361.} The following introductory remarks in
Einstein's dissertation indicate what he adds to this.
\begin{quote}
``\textit{The earliest determinations of real sizes of molecules were possible by the
kinetic theory of gases, whereas the physical phenomena observed in liquids have thus far
not served for the determination of molecular sizes. This is no doubt due to the fact that
it has not yet been possible to overcome the obstacles that impede the development of a
detailed molecular-kinetic theory of liquids. It will be shown in this paper that the size
of molecules of substances dissolved in an undissociated dilute solution can be obtained
from the internal friction of the solution and the pure solvent, and from the diffusion of
the dissolved substance within the solvent. (...).}''
\end{quote}

Beside originality and intuition, great scientists usually also dispose of a fair amount of
technical abilities. That Einstein was not an exception in this respect, should become
clear if we now go into the technical details of his dissertation.

\subsection{Basic equations of hydrodynamics}

Let me first recall some general facts of hydrodynamics, that we shall need. I will use
notation that has become standard, and not the one that was common at the time when
Einstein did his work.

For stationary incompressible flows of homogeneous fluids, the Navier-Stokes equation is
\[ (\mathbf{v}\cdot\nabla)\mathbf{v}=-\frac{1}{\rho}\nabla
p+\frac{\eta}{\rho}\triangle\mathbf{v}.\] We consider only situations with small Reynold
numbers.Then one can neglect the left-hand side, and the basic equations become
\begin{equation} \nabla
p=-\eta\triangle\mathbf{v}, ~~~\nabla\cdot\mathbf{v}=0.
\end{equation}
These imply that the pressure is harmonic: $\triangle p=0$. The same holds for the
vorticity $\mathrm{curl}~\mathbf{v}$. We also recall the expression for the stress tensor
\begin{equation}
\sigma_{ij}=-p\delta_{ij}+\eta(\partial_iv_j+\partial_jv_i).
\end{equation}
According to (1) this is divergence-free: $\partial_j\sigma_{ij}=0$. Later we shall also
need the following expression for the rate $W$ at which the stresses do work on the surface
$\partial\Omega$ bounding a region $\Omega$:
\begin{equation}
W=\int_{\partial\Omega}v_i\sigma_{ij}n_j~dA.
\end{equation}
Here, $\mathbf{n}$ is the outward pointing unit vector.

\subsection{Einstein's strategy}

With Einstein we now consider an incompressible fluid of viscosity $\eta_0$, in which a
large number of identical, rigid, spherical particles is inserted. This suspension can be
described in two ways: (1) On large scales, in comparison to the average separation of
neighboring solute particles, as a homogeneous medium with an effective viscosity $\eta$.
(2) By the stationary flow of the fluid (solvent) that is modified by the suspended
particles.

For both descriptions Einstein computes according to (3) the rate of work for a big region
$\Omega$ and obtains by equating the two results the important formula
\begin{equation}
\eta=\eta_0\left(1+\frac{5}{2}\varphi\right),
\end{equation}
where $\varphi$ denotes the fraction of the volume occupied by the suspended particles.
This is assumed to be small (dilute suspension). (Due to a calculational error, Einstein
originally lost the factor 5/2; we shall come back to this amusing story.)

\subsection{Velocity field for a single suspended particle}

We first adopt the second description. As a preparing task we have to determine the
modification of a flow with constant velocity gradient, say, caused by a single little
ball. Mathematically, we have to solve a boundary value problem for the elliptic system
(1).

So let the unperturbed velocity field be
\begin{equation}
v_i^{(0)}=e_{ij}x_j,
\end{equation}
where $e_{ij}$ is a constant, symmetric, traceless tensor. The last property reflects the
incompressibility. $e_{ij}$ is the deformation tensor (we are not interested in flows with
non-vanishing vorticity). The unperturbed pressure is denoted by $p^{(0)}$. The stress
tensor for the background flow is
\begin{equation}
\sigma_{ij}^{(0)}=-p^{(0)}\delta_{ij}+2\eta_0e_{ij}.
\end{equation}

We decompose the modified velocity field $\mathbf{v}$ according to
\begin{equation}
\mathbf{v}=\mathbf{v}^{(0)}+\mathbf{v}^{(1)}
\end{equation}
into an unperturbed part plus a perturbation $\mathbf{v}^{(1)}$. The boundary conditions
are: $\mathbf{v}=0$ on the ball with radius $a$ and $\mathbf{v}=\mathbf{v}^{(0)}$ at
infinity. Analogous decompositions are used for the pressure and the stresses:
\begin{equation}
p=p^{(0)}+p^{(1)}, ~~~\sigma_{ij}=\sigma_{ij}^{(0)}+\sigma_{ij}^{(1)},
\end{equation}
where
\begin{equation}
\sigma_{ij}^{(1)}=-p^{(1)}\delta_{ij}+\eta_0(\partial_iv_j^{(1)}+\partial_jv_i^{(1)}).
\end{equation}
For $W$ we then have the decomposition ($|\Omega|$= volume of $\Omega$)
\begin{equation}
W=2\eta_0e_{ij}e_{ij}|\Omega|+e_{ik}\int_{\partial\Omega}\sigma_{ij}^{(1)}x_kn_j~dA
+\int_{\partial\Omega}v_i^{(1)}\sigma_{ij}^{(0)}n_j ~dA.
\end{equation}

The rigid ball is taken as the origin of a cartesian coordinate system. Einstein determines
the perturbations  $v_i^{(1)}$ and $p^{(1)}$ with the help of a method which is described
in Kirchhoff's ``Vorlesungen \"{u}ber Mechanik''\footnote{G. Kirchhoff, \textit{Vorlesungen
\"{u}ber mathematische Physik}, Vol. 1, \textit{Mechanik}, Teubner (1897).}, which he had
studied during his student years. This involves the following two steps: a) Determine a
function $V$, which satisfies the equation
\begin{equation}
\triangle V=\frac{1}{\eta_0}p^{(1)},
\end{equation}
and set
\begin{equation}
v_i^{(1)}=\partial_iV+v_i',
\end{equation}
where $v_i'$ has to satisfy the following equations
\begin{equation}
\triangle v_i'=0, ~~~\partial_iv_i'=-\frac{1}{\eta_0}p^{(1)}.
\end{equation}
\textit{Remark} on a). As a consequence of (11)-(13) the basic equations for $v_i^{(1)}$
and $p^{(1)}$ are satisfied:
\[ \eta_0\triangle v_i^{(1)}=\eta_0\partial_i\triangle V=\partial_ip^{(1)}, ~~
\partial_iv_i^{(1)}=\triangle V+\partial_iv_i'=0.\]
b) Use the following decaying harmonic ansatz for $p^{(1)}$
\begin{equation}
\frac{p^{(1)}}{\eta_0}=Ae_{ij}\partial_i\partial_j\left(\frac{1}{r}\right)
\end{equation}
with a constants  $A$, and try for $v_i'$ the harmonic expression
\begin{equation}
v_i'=-\tilde{A}e_{ik}\partial_k\left(\frac{1}{r}\right)
+B\partial_ie_{jk}\partial_j\partial_k\left(\frac{1}{r}\right).
\end{equation}
This fulfills  both equations (13) for $\tilde{A}=A$, because we then have
\[\partial_iv_i'=-Ae_{ik}\partial_i\partial_k\left(\frac{1}{r}\right)=-\frac{p^{(1)}}{\eta_0}.\]
As a result of $\triangle r=2/r$, equation (11) is satisfied for
\begin{equation}
V=\frac{1}{2}Ae_{ij}\partial_i\partial_j r.
\end{equation}
Performing the differentiations in (15) and (16), we obtain

\begin{equation}
v_i^{(1)}=\frac{3}{2}Ae_{jk}\frac{x_ix_jx_k}{r^5}+
B\left(6e_{ik}\frac{x_k}{r^5}-15e_{jk}\frac{x_ix_jx_k}{r^7}\right).
\end{equation}
The boundary condition $v_i^{(1)}=-e_{ij}x_j$ for $r=a$ requires
\begin{equation}
A=-\frac{5}{3}a^3, ~~~B=-\frac{a^5}{6}.
\end{equation}
We thus obtain for the perturbation of the velocity field ($n_i:=x_i/r$)
\begin{equation}
v_i^{(1)}=-\frac{5}{2}a^3e_{jk}\frac{1}{r^2}n_in_jn_k-
\frac{a^5}{6}\left(6e_{ik}\frac{x_k}{r^5}-15e_{jk}\frac{x_ix_jx_k}{r^7}\right).
\end{equation}
According to (12) and (15) for $\tilde{A}=A$ we can represent $v_i^{(1)}$ also as follows
\begin{equation}
v_i^{(1)}=-\frac{5}{6}a^3e_{jk}\partial_i\partial_j\partial_k (r)
+\frac{5}{3}a^3e_{ik}\partial_k\left(\frac{1}{r}\right)
-\frac{1}{6}a^5\partial_ie_{jk}\partial_j\partial_k\left(\frac{1}{r}\right).
\end{equation}
Equation (14) gives for the pressure
\begin{equation}
p=p^{(0)}-5\eta_0a^3e_{ij}\frac{n_in_j}{r^3}.
\end{equation}

Einstein claims that it can be demonstrated that his solution of the boundary value problem
is unique, but he gives only some indications of what he thinks is a proof. Apparently, he
did not know that an elegant uniqueness proof for such problems was already given  in 1868
by Helmholtz.\footnote{H. Helmholtz, ``Theorie der station\"{a}ren Str\"{o}me in reibenden
Fl\"{u}ssigkeiten'', Wiss. Abh., Bd. I, S. 223.} Consider for two solutions of the basic
equations, for given velocity fields on the boundaries, the non-negative quantity
$(\theta'_{ij}-\theta_{ij})(\theta'_{ij}-\theta_{ij})$, where $\theta'_{ij},\theta_{ij}$
are the deformation tensors of the two velocity fields. It is easy to show that the
integral of this function over the region outside the bodies must vanish. (Use partial
integrations and the basic equations (1).) Therefore, $\theta'_{ij}=\theta_{ij}$. In other
words, the deformation tensor for the difference $v'_i-v_i$ of the two velocity fields
vanishes. This difference is thus a combination of a rigid translation and a rigid
rotation. Because of the imposed boundary conditions, the two velocity fields must agree.
The pressures for the two solutions are, therefore, also the same, up to an additive
constant.

\subsection{Two expressions for the rate $W$}

In (10) we now choose for $\Omega$ a large ball $K_R$ with radius $R$. In leading order
only the first term of (19) contributes, and a routine calculation leads to the following
expression for $W$ in terms of the spherical moments
\[
\overline{n_in_jn_kn_l}:=\frac{1}{4\pi}\int_{S^2}n_in_jn_kn_l
~d\Omega=\frac{1}{15}(\delta_{ij}\delta_{kl}+
\delta_{ik}\delta_{jl}+\delta_{il}\delta_{lk}),\]
\begin{eqnarray*}
W &=& 2\eta_0e_{ij}e_{ij}|\Omega|+20\pi
a^3\eta_0e_{ik}\{3e_{rs}\overline{n_in_kn_rn_s}-e_{is}\overline{n_sn_k}\} \\
&=& 2\eta_0e_{ij}e_{ij}\left\{|\Omega|+\frac{1}{2}\frac{4\pi}{3}a^3\right\}.
\end{eqnarray*}
This holds for a single ball. As long as the suspension is dilute, we obtain Einstein's
(corrected) result
\begin{equation}
W=2\eta_0e_{ij}e_{ij}|\Omega|\left(1+\frac{1}{2}\varphi\right).
\end{equation}

Following Einstein, we now calculate the same quantity by adopting the first description of
the suspension.  For this we write the result (20) in the form
\begin{equation}
v_i=e_{ij}x_j+(e_{ik}\triangle-e_{jk}\partial_i\partial_j)\partial_kf,
\end{equation}
with
\begin{equation}
f=-\frac{1}{2}Ar-\frac{B}{r}
\end{equation}
($A$ and $B$ have the earlier meaning (18)). When the contributions of all the suspended
balls, with number density $n$, inside the ball $K_R$ are summed, we obtain for the
velocity field in $K_R$
\begin{equation}
v_i=e_{ij}x_j+(e_{ik}\triangle-e_{jk}\partial_i\partial_j)\partial_kF,
\end{equation}
where
\begin{equation}
F(|\mathbf{x}|)=n\int_{K_R}f(|\mathbf{x}-\mathbf{x}'|)~d^3x'=\frac{\pi}{3}nA
\left(\frac{1}{10}r^4-r^2R^2\right) -2\pi n B\left(R^2-\frac{1}{3}r^2\right).
\end{equation}
From this one easily finds
\begin{equation} v_i=e_{ij}x_j(1-\varphi).
\end{equation}
Einstein obtained this result slightly differently. We can use it to obtain a second
expression for $W$:
\begin{equation}
W=2\eta e_{ij}e_{ij}|\Omega|(1-2\varphi).
\end{equation}
Comparing (22) and (28) leads to the announced formula (4) due to Einstein.

\subsection{Two relations between the Avogadro number \\ and the molecular radius.}

If the rigid balls are molecules, for instance sugar, then
\begin{equation}
\varphi=\frac{4\pi}{3}a^3\frac{N_A\rho_s}{m_s},
\end{equation}
where $\rho_s$ is the mass density of the solute and $m_s$ its molecular weight which were
known to Einstein. In addition, there existed measurements of $\eta/\eta_0$ for dilute
sugar solutions. Hence, Einstein obtained  from (4) a relation between  $N_A$ and $a$.

Using available data, Einstein states the following.\footnote{I give here the later numbers
from 1911.} ``One gram of sugar dissolved in water has the same effect on the coefficient
of viscosity as do small suspended rigid spheres of a total volume of $0.98~cm^3$.'' On the
other hand, the \textit{density} of an aqueous sugar solution behaves experimentally as a
mixture of water and sugar in dissolved form with a specific volume of $0.61~cm^3$. (The
latter is also the volume of one gram of solid sugar.) Einstein interprets the difference
of the two numbers as due to an attachment of water molecules to each sugar molecule. The
radius $a$ in (29) is thus a ``hydrodynamically effective radius'' of the molecule, which
takes the enlargement due to hydration into account.
\subsubsection*{Diffusion}

In order to be able to determine the two quantities individually, Einstein searched for a
second connection, and thereby found his famous diffusion formula. Its derivation is quite
short, but ``extremely ingenious'' (A. Pais). It rests on thermal and mechanical
equilibrium considerations.

Assume that a constant external force $\mathbf{f}$  acts on the the suspended particles.
This causes a particle current of magnitude  $n\mathbf{v}$, where $n$ is the number density
and $\mathbf{v}=$ the velocity of the particle current. In equilibrium this is balanced by
the diffusion current $-D\nabla n,~D=$ diffusion constant. The velocity of the particle
current is proportional to $\mathbf{f}$,
\begin{equation}
\mathbf{v}=b\mathbf{f},~~~b:\mathnormal{mobility}.
\end{equation}
These considerations give us the (dynamical) equilibrium condition
\begin{equation}
D\nabla n=nb\mathbf{f}.
\end{equation}

In thermal equilibrium, the external force is balanced by the gradient of the osmotic
pressure. According to the law of van't Hoff\footnote{According to this, the osmotic
pressure $p$ exerted by the suspended particles is exactly the same as if they alone were
present as an ideal gas. In equilibrium we thus have $n\mathbf{f}=\nabla p=kT\nabla n$.}
this means
\begin{equation}
\mathbf{f}=\frac{kT}{n}\nabla n.
\end{equation}
Inserting this into the last relation leads to the simple formula
\begin{equation}
D=kTb.
\end{equation}
For the mobility Einstein uses Stokes' relation
\begin{equation}
b=\frac{1}{6\pi\eta_0a}
\end{equation}
and obtains in this way his famous formula
\begin{equation}
D=\frac{kT}{6\pi\eta_0a},~~k=\frac{R}{N_A}
\end{equation}
($R=$ gas constant).

This was almost simultaneously discovered in Australia by  William Sutherland.

A beauty of the argument is that the exterior force drops out. Similar equilibrium
considerations between systematic and fluctuating forces were repeatedly made by Einstein.

\subsection{Silence, a calculational error, late attention}

By 1909 Perrin's careful measurements of Brownian motion led to a new value for
Avogadro's number that was significantly different from the value Einstein had obtained
from his thesis work, and also somewhat different from what he and Planck had deduced from
black-body radiation. Einstein then drew Perrin's attention to his hydrodynamical method,
and suggested its application to the suspensions studied by Perrin. Then Jacques Bancelin,
a Pupil of Jean Perrin, checked Einstein's viscosity formula $\eta=\eta_0(1+\varphi)$.
Bancelin confirmed that there was an increase of the viscosity that was independent of the
size of the suspended particles, and only depends on the total volume they occupy. However,
he got a stronger increase. Initially, this increase was too steep; in the publication
Bancelin gives the result $\eta=\eta_0(1+2.9\varphi)$.

On 27 December, 1910 Einstein wrote from Z\"{u}rich to his former student and collaborator
Ludwig Hopf about the puzzling situation, and then adds:
\begin{quote}
``\textit{I have checked my previous calculations and arguments and found no error in them.
You would be doing a great service in this matter if you would carefully recheck my
investigation. Either there is an error in the work, or the volume of Perrin's suspended
substance in the suspended state is greater than Perrin believes.}''
\end{quote}

Hopf indeed found an error in some differentiation process, and got the formula (4).
Einstein communicated the result to Perrin, and published in (1911) a correction of his
thesis in the \textit{Annalen}. (By the way, this correction is the second most quoted
paper of Einstein.) New experimental data for sugar solutions now gave the excellent value
\begin {equation}
N_A=6.56\times10^{23}
\end{equation}
for the  Avogadro number, in good agreement with the results of other methods, in
particular with Perrin's determination from the Brownian motion, for which he got the Nobel
price in 1926. Both results were discussed by Perrin in his extensive report at the famous
Solvay conference in 1911.

\section{Final remarks}

In his `Autobiographical Notes' of 1949, what he called his `necrology', Einstein only
briefly describes his applications of classical statistical mechanics. The thesis is not
mentioned at all. About the law of Brownian motion he says:
\begin{quote}
``\textit{The agreements of these considerations with experience together with Planck's
determination of the true molecular size from the law of radiation (for high temperatures)
convinced the sceptics, who were quite numerous at the time (Ostwald, Mach) of the reality
of atoms. The antipathy of these scholars toward atomic theory can indubitably be traced
back to their positivistic philosophical attitude. This is an interesting example of the
fact that even scholars of audacious spirit and fine instinct can be obstructed in the
interpretation of facts by philosophical prejudices.}''
\end{quote}

Perrin's famous book ``Les Atomes'' of 1913, a classic of twentieth century
physics\footnote{A new edition of the original text has appeared in Flammarion (1991), ISBN
2-08-081225-4; for an English translation, see: J. Perrin, \textit{Atoms}, Van Nostrand
(1916).}, ends with the words:
\begin{quote}
``\textit{The atomic theory has triumphed. Until recently still numerous, its adversaries,
at last overcome, now renounce one after another their misgivings, which were, for so long,
both legitimate and undeniably useful.}''
\end{quote}

Einstein's very decent value (36) is not quoted in Perrin's book. This indicates that
Einstein's thesis was not widely appreciated in the early years. For this reason Einstein
published in 1920 a brief note, drawing attention to his erratum from 1911, ``which till
now seems to have escaped the attention of all who work in this field''.

Since Einstein was so fond of applying physics to practical situations, he would certainly
have enjoyed hearing that his doctoral thesis found so many applications.
\end{document}